\documentclass{article}
\usepackage{graphicx}
\usepackage{ai4e}
\usepackage{subcaption}

\usepackage[nameinlink]{cleveref}
\crefname{figure}{Fig.}{Figs.}
\Crefname{figure}{Figure}{Figures}
\crefname{equation}{Eq.}{Eqs.}
\Crefname{equation}{Equation}{Equations}
\crefformat{section}{#2§#1#3}
\crefname{section}{§}{§§}
\Crefname{section}{Section}{Sections}
\crefname{table}{Table}{Tables}
\crefname{appendix}{Appendix}{Appendices}

\newcommand{\ie}{i.e.,\ }

\title{HomeLabGym: A real-world testbed for \\
home energy management systems}

\author{
  Toon Van Puyvelde, \thanks{Under Review} \\
  IDLab \\
  Ghent University -- imec \\
  \texttt{toon.vanpuyvelde@ugent.be} \\
  \And
  Marie-Sophie Verwee \\
  IDLab \\
  Ghent University -- imec \\
   \And
  Gargya Gokhale \\
  IDLab \\
  Ghent University -- imec \\
  \And
  Mehran Zareh Eshghdoust \\
  Daikin Europe NV \\
   \And
  Chris Develder \\
  IDLab \\
  Ghent University -- imec \\
}

\begin{document}
\maketitle

\begin{abstract}
Amid growing environmental concerns and resulting energy costs, there is a rising need for efficient Home Energy Management Systems (HEMS).
Evaluating such innovative HEMS solutions typically relies on simulations that may not model the full complexity of a real-world scenario.
On the other hand, real-world testing, while more accurate, is labor-intensive, particularly when dealing with diverse assets, each using a distinct communication protocol or API.
Centralizing and synchronizing the control of such a heterogeneous pool of assets thus poses a significant challenge.
In this paper, we introduce HomeLabGym, a real-world testbed to ease such real-world evaluations of HEMS and flexible assets control in general, by adhering to the well-known OpenAI Gym paradigm.
HomeLabGym allows researchers to prototype, deploy, and analyze HEMS controllers within the controlled test environment of a real-world house (the IDLab HomeLab), providing access to all its available sensors and smart appliances.
The easy-to-use Python interface eliminates concerns about intricate communication protocols associated with sensors and appliances, streamlining the evaluation of various control strategies.  
We present an overview of HomeLabGym, and demonstrate its usefulness to researchers in a comparison between real-world and simulated environments in controlling a residential battery in response to real-time prices.
\end{abstract}

\section{Introduction}
The surge in energy costs, coupled with the growing emphasis on demand-response and integration of renewable energy sources (RES), has increased interest in the use of home energy management systems (HEMS).
Effective HEMS can strategically schedule loads such as heating, ventilation, and air conditioning (HVAC), and electric vehicle (EV) charging, while optimizing the usage of local energy sources, e.g., photovoltaic (PV) installations and Energy Storage Systems (ESS)~\cite{WANG2020115036}.
HEMS control algorithms are typically tested in simulation environments, notable examples being, e.g., ModelicaGym~\cite{modelicagym} for integrating Functional Mockup Units (FMUs), CityLearn~\cite{citylearn} for grid-level energy distribution, and BOPTEST for building control~\cite{boptest}.
Yet, no easily reproducible \emph{real-world} testbed has been widely adopted for real-world deployment of (residential) flexible asset controllers.
Since realistic simulation of real-world households is both complex to model and to run, we believe there is a need for such easy-to-use real-world test environments to deploy, validate, and analyze flexible asset controllers.
Inspired by real-world testbeds in other domains --- e.g., OffWorld Gym  for robotics~\cite{kumar2020offworld} --- we thus propose to use a real-world environment for plug-and-play testing of HEMS controllers.
In particular, we use the HomeLab,\footnote{{https://homelab.ilabt.imec.be/}} a state-of-the-art smart home for realistic experimentation developed by the IDLab research group at Ghent University, Belgium.
However, accessing the broad set of sensors and controlling the wide range of devices (e.g., heat pumps, battery chargers, PV inverters) therein, requires a diverse set of protocols/interfaces.
To abstract away this complexity for HEMS control researchers, we created HomeLabGym, a real-world testbed that is customizable to specific use cases of researchers. 

\begin{figure}[ht]
  \centering
  \includegraphics[width=\linewidth]{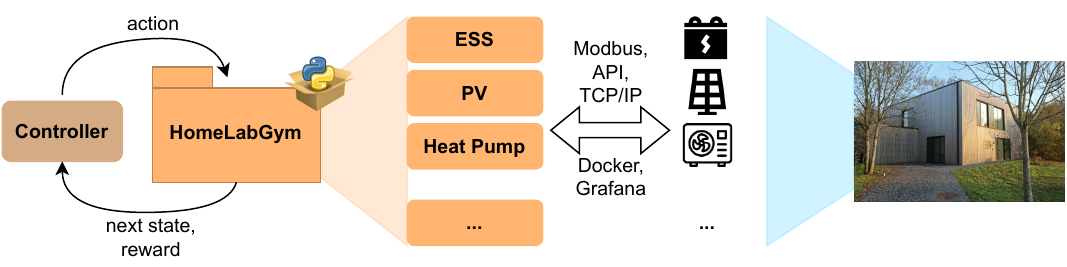} 
  \caption{System Architecture of HomeLabGym.}
  \label{fig:structure}
\end{figure}

\section{HomeLabGym}
The main objective of HomeLabGym is to offer an easy-to-use environment for real-world (or at least realistic) deployment of innovative HEMS controllers, particularly supporting controllers based on state-of-the-art data-driven reinforcement learning (RL).
In that particular field, OpenAI Gym is a common platform for testing RL innovations with minimal coding efforts~\cite{towers_gymnasium_2023}.
As illustrated in \cref{fig:structure}, the Python based HomeLabGym package incorporates a modular design that interfaces with different assets of the real-world testbed \ie IDLAB HomeLab. 

Each asset has a corresponding module for implementing the necessary communication protocols, e.g., MODBUS for the ESS and ENTSOE-API for the day-ahead price. HomeLabGym also includes additional functionalities such as preprocessing (for register-level conversions), data logging, Grafana-based visualization and support for containerizing and cluster-based deployments. 
In our first version, the HomeLabGym integrates existing HomeLab assets which include the heat pump and thermostat, the home battery system, solar PV system and smart energy meters.
Additionally, the current version includes the day-ahead energy prices for the BELPEX region.
To warrant adaptability of HomeLabGym to future extensions of the physical HomeLab infrastructure, e.g., an EV charger, the package is built in a modular manner as illustrated in \cref{fig:structure}.

\section{Experiment}
The objective of the short experiment is to illustrate the ``plug-and-play'' nature of our HomeLabGym solution, in terms of using standard implementations of control algorithms, e.g., based on reinforcement learning.
Additionally, we demonstrate the usefulness of such real-world experiments in uncovering complexities that are difficult to model/reveal purely in simulation.

The case study we consider is of using a home battery to energy arbitrage based on a real-time price profile using BELPEX spot prices.\footnote{This refers to the organised wholesale market for power trading in the Belgium energy market, \ie the current European Power Exchange Belgium.}
The control algorithm we use is a Deep Q-Network (DQN) agent, using the implementation from the Stable Baselines3 RL library,\footnote{https://stable-baselines3.readthedocs.io/en/master/modules/dqn.html} out-of-the-box, as is, given our HomeLabGym's compatibility with OpenAI Gym.
The DQN agent is trained using 2023 Belpex prices in simulation.
We then simply deploy the trained DQN model in our HomeLab, using HomeLabGym, for a 4-day testing period. 
As both the battery simulator and HomeLabGym are Gymnasium environments, this was a plug-and-play transfer, which not only accelerates the deployment process but also showcases the versatility and ease of integration offered by HomeLabGym.

\begin{figure}
    \centering
    \includegraphics[width=\linewidth]{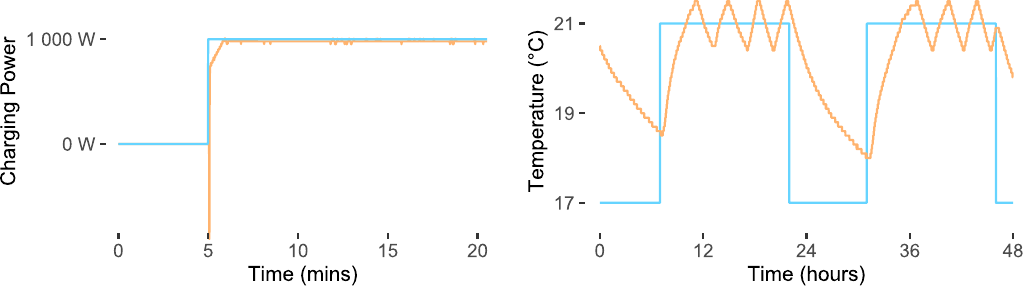}
    \begin{subfigure}{0.48\linewidth}
        \centering
        \caption{Charging power of the battery.}
        \label{fig:power_plot}
    \end{subfigure}
    \hfill
    \begin{subfigure}{0.48\linewidth}
        \centering
        \caption{2-day room temperature of the living room.}
        \label{fig:temp_plot}
    \end{subfigure}
    \caption{Graphs of real-world assets in HomeLab.}
    \label{fig:experiment}
\end{figure}

As an example of added value that \emph{real-world} testing can provide, we study the daily reward achieved with the considered DQN agent.
Particularly, we compare the cumulative reward attained with the one a purely simulation-based experiment (with the same DQN agent) would have attained.
Our experiment revealed that the real-world reward is
2\% lower than that attained in simulation. 
This disparity can be attributed to variations between real-world conditions and simulation environments. 
Specifically, \cref{fig:power_plot} illustrates the variability in actual charging power, often falling below the 1\,kW setpoint, resulting in a non-linear charging speed. 
Similarly, \cref{fig:temp_plot} illustrates the (step-like) behavior of the heat pump as it approaches the room heating set-point, giving us more insight into the hardware-level discrepancies of the heat pump. 

\section{Conclusion}
HEMS are often tested and benchmarked on simulated environments. 
However, deploying controllers in real-world households requires additional coding and an in-depth understanding of various communication protocols to interface with heterogeneous assets (e.g., batteries, HVAC systems, EV chargers). 
Also, as our experiment shows, simulations do not fully align with real-world dynamics.
To allow easy prototyping of flexible asset controllers on a controlled real-world household, we have introduced HomeLabGym, a Python interface to easily test controllers in a real-world home (i.e., the IDLab HomeLab). 
Our HomeLabGym environment supports a wide range of communication protocols, thus providing a plug-and-play tool to help investigate the real-world performance of HEMS controllers and analyze their deployability. 
The IDLab HomeLab is open for collaborations across the world to access the HomeLab testbed facilities.

\section*{Acknowledgement}
This research was funded by Daikin Europe NV under the PhD Framework Agreement with Ghent University. For more information on HomeLab, please contact matthias.strobbe@ugent.be.

\bibliographystyle{unsrt}  
\bibliography{references} 

\begin{thebibliography}{1}

\bibitem{WANG2020115036}
Zhe Wang and Tianzhen Hong.
\newblock Reinforcement learning for building controls: The opportunities and challenges.
\newblock {\em Applied Energy}, 269:115036, 2020.

\bibitem{modelicagym}
Oleh Lukianykhin and Tetiana Bogodorova.
\newblock Modelicagym: applying reinforcement learning to modelica models.
\newblock In {\em Proceedings of the 9th International Workshop on Equation-based Object-oriented Modeling Languages and Tools}, EOOLT ’19. ACM, November 2019.

\bibitem{citylearn}
Jose~R Vazquez-Canteli, Sourav Dey, Gregor Henze, and Zoltan Nagy.
\newblock Citylearn: Standardizing research in multi-agent reinforcement learning for demand response and urban energy management, 2020.

\bibitem{boptest}
David Blum, Javier Arroyo, Sen Huang, J{\'a}n Drgo{\v{n}}a, Filip Jorissen, Harald~Taxt Walnum, Yan Chen, Kyle Benne, Draguna Vrabie, Michael Wetter, et~al.
\newblock Building optimization testing framework (boptest) for simulation-based benchmarking of control strategies in buildings.
\newblock {\em Journal of Building Performance Simulation}, 14(5):586--610, 2021.

\bibitem{kumar2020offworld}
Ashish Kumar, Toby Buckley, John~B. Lanier, Qiaozhi Wang, Alicia Kavelaars, and Ilya Kuzovkin.
\newblock Offworld gym: open-access physical robotics environment for real-world reinforcement learning benchmark and research, 2020.

\bibitem{towers_gymnasium_2023}
Mark Towers, Jordan~K. Terry, Ariel Kwiatkowski, John~U. Balis, Gianluca~de Cola, Tristan Deleu, Manuel Goulão, Andreas Kallinteris, Arjun KG, Markus Krimmel, Rodrigo Perez-Vicente, Andrea Pierré, Sander Schulhoff, Jun~Jet Tai, Andrew Tan~Jin Shen, and Omar~G. Younis.
\newblock Gymnasium, March 2023.

\end{thebibliography}
\end{document}